\documentclass[conference]{IEEEtran}

\usepackage{amsmath,amssymb}
\usepackage{algorithm}
\usepackage{algorithmic}
\usepackage{booktabs}
\usepackage{cite}
\usepackage{graphicx}
\usepackage{microtype}
\usepackage{url}
\graphicspath{{figures/}}

\newcommand{\MIPAllFeasible}{83.3\%}
\newcommand{\MIPAllOptimal}{33.3\%}

\newcommand{\CoupledAllFeasible}{6/6}
\newcommand{\CoupledTwentySevenFeasible}{3/3}
\newcommand{\CoupledTwentySevenCost}{2522.499}
\newcommand{\CoupledTwentySevenGap}{87.77\%}
\newcommand{\CoupledTwentySevenEnergy}{138.515}

\newcommand{\CoupledTwentySevenLatency}{84.84}

\newcommand{\CoupledMoEFeasible}{3/3}
\newcommand{\CoupledMoECost}{1592.697}
\newcommand{\CoupledMoEGap}{18.56\%}
\newcommand{\CoupledMoEEnergy}{104.439}
\newcommand{\CoupledMoETempRange}{23.6001--23.8000}
\newcommand{\CoupledMoELatency}{33.92}

\newcommand{\CoupledGurobiCost}{1343.380}
\newcommand{\CoupledGurobiEnergy}{92.052}
\newcommand{\CoupledGurobiMedian}{0.0487}
\newcommand{\CoupledHVACEnergyAfterWindow}{0.000}
\newcommand{\CoupledTwentySevenVsMoECostDelta}{929.802}
\newcommand{\CoupledTwentySevenVsMoECostPct}{58.38\%}

\begin{document}

\title{Physically Constrained Agentic AI for Energy Scheduling}

\author{\IEEEauthorblockN{Dafang Zhao\IEEEauthorrefmark{1}\IEEEauthorrefmark{2}, Yang Deng\IEEEauthorrefmark{3}, Zhengmao Li\IEEEauthorrefmark{4}, Zekun Guo\IEEEauthorrefmark{5}, Ka Tai Lau\IEEEauthorrefmark{6}, and Samson Tai\IEEEauthorrefmark{7}}
\IEEEauthorblockA{\IEEEauthorrefmark{1}\textit{The University of Osaka, Japan},
\IEEEauthorrefmark{2}\textit{Kinevo Limited, Hong Kong}\\
\IEEEauthorrefmark{3}\textit{Saint Francis University, Hong Kong},
\IEEEauthorrefmark{4}\textit{Aalto University, Finland}\\
\IEEEauthorrefmark{5}\textit{University of Hull, United Kingdom}\\
\IEEEauthorrefmark{6}\textit{Electrical and Mechanical Services Department, HKSAR Government, Hong Kong}\\
\IEEEauthorrefmark{7}\textit{Hong Kong Baptist University, Hong Kong}\\}}

\maketitle

\begin{abstract}
Agentic AI extends energy management beyond fixed-form interaction by
translating natural-language requests into coordinated scheduling actions.
We present a hierarchical ReAct Energy Management System (EMS) in which one orchestrator coordinates
LLM-based specialist roles for shiftable appliances, EV charging, and thermal
control. Physical authorization remains strictly separate from
language-mediated schedule generation in this framework. A deterministic critic
reconstructs each integrated day-ahead candidate and checks its schema,
appliance cycles, device power, and thermal comfort. When active, it also
enforces the whole-home feeder limit. Across Qwen~3.5 checkpoints,
single-appliance mixed-integer schedules were feasible in \MIPAllFeasible\ of
runs. Localized feedback produced no accepted coupled schedule, whereas a
multi-step policy authorized \CoupledAllFeasible\ current coupled runs:
\CoupledTwentySevenFeasible\ for 27B and \CoupledMoEFeasible\ for 35B-A3B.
The standard occupied-window policy permits pre-conditioning and enforces comfort
from 09:00--18:00. Every accepted schedule then passed an independent final
replay without modification. Feasible costs were
\CoupledTwentySevenCost~JPY for 27B and \CoupledMoECost~JPY for 35B-A3B,
slightly above the mathematical-programming optimum of \CoupledGurobiCost~JPY. These results establish a
fail-closed workflow for agentic MIP and MILP energy scheduling under the
declared physical model.
\end{abstract}

\begin{IEEEkeywords}
agentic AI, energy management, large language models, mixed-integer
linear programming, physical constraint critic
\end{IEEEkeywords}

\section{Introduction}

Day-ahead household energy scheduling must reconcile user intent with tariffs,
weather, calendar constraints, and heterogeneous device models. Large language
model (LLM) interfaces can translate dialogue into typed scheduling parameters
\cite{michelon2025interface}. Agentic AI extends this interface by allowing an
LLM to plan, invoke tools, observe outcomes, and revise actions. The ReAct
pattern operationalizes this loop by interleaving language-based reasoning with
task-specific actions \cite{yao2023react}. Recent energy-management systems use
this capability to delegate heterogeneous decisions and coordinate
device-specific schedules \cite{elmakroum2026agentic,jung2026hema}.

The word \emph{agent} also has an established meaning in agent-based modeling,
where it denotes an autonomous model entity with specified states, behaviors,
and interaction rules \cite{bonabeau2002agent}. In this study, an appliance
specialist instead denotes a prompt-defined scheduling role executed by an LLM
within a ReAct workflow. It interprets a delegated subproblem, returns a typed action proposal,
and revises that proposal after critic feedback. These specialists neither
simulate appliances as model entities nor independently authorize physical
control. We therefore use \emph{LLM-based specialist} for these roles and
reserve \emph{agent-based model} for the conventional modeling paradigm.

Language-mediated coordination is useful, but a plausible schedule does not
establish feasibility. Individually reasonable specialist outputs can conflict
after integration because appliances share a feeder and heat-pump decisions
alter a common thermal state. Simulator feedback and physics-informed agent
loops provide mechanisms for detecting such conflicts
\cite{jia2025feedback,jiang2026physics}. However, feedback alone does not define
which exact candidate may cross the control boundary. The central challenge is
therefore authorization: flexible semantic reasoning must remain subordinate to
deterministic evaluation of the encoded physical constraints.

We address this challenge by separating schedule generation from physical
authorization. A hierarchical ReAct orchestrator coordinates specialist
proposals, while a deterministic critic alone may release the final integrated
candidate. This paper makes three contributions, beginning with a mapping from
natural-language requests to coordinated LLM-based specialist actions. Second,
it defines a fail-closed critic with sign-safe thermal feedback, coupled
multi-step revision, and explicit rejection. Third, it evaluates discrete,
continuous, and coupled scheduling across six Qwen~3.5 checkpoints with matched
mathematical-programming references. Accordingly, the evidence supports
feasibility only under the encoded physical model.

\section{Agentic EMS and Physical Authorization}

\begin{figure}[t]
  \centering
  \includegraphics[width=\columnwidth]{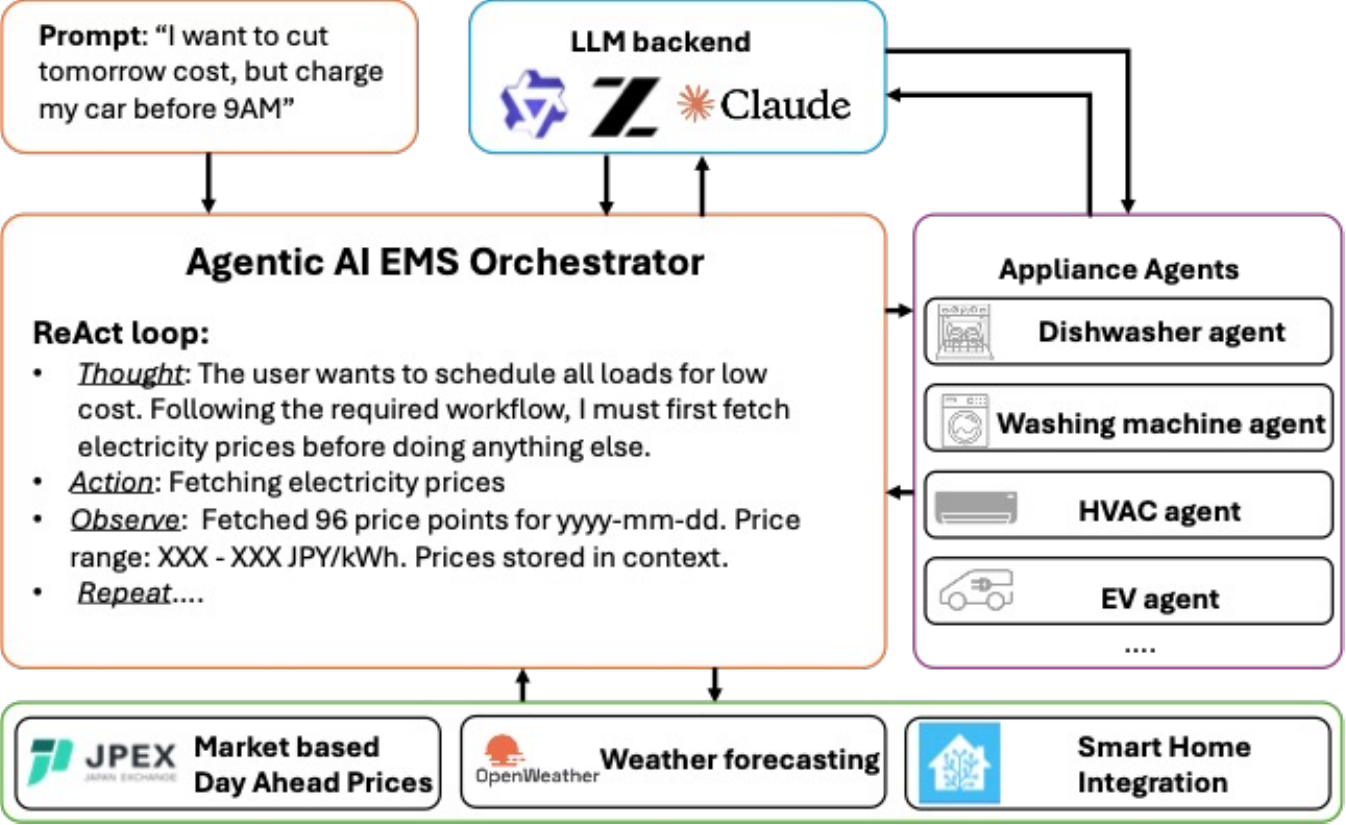}
  \caption{Agentic-AI EMS overview. The ReAct orchestrator combines a
  household request with shared market and weather context and coordinates
  appliance specialists through a configurable LLM backend. The ``Appliance
  Agents'' block denotes LLM-based specialist roles, not agent-based model
  entities.}
  \label{fig:overview-short}
\end{figure}

\subsection{Hierarchical ReAct coordination}

The orchestrator observes the request, selects actions, obtains one frozen
tariff and weather context, delegates typed subproblems, and integrates the
specialist observations (Fig.~\ref{fig:overview-short}). Three LLM-based
specialist roles cover shiftable appliances, EV charging, and heat-pump control;
the shiftable-appliance role is instantiated by separate washing-machine and
dishwasher prompts. The discrete-scheduling specialists return one integer start for an
uninterrupted cycle, while the thermal-control specialist returns a variable-power profile.
The integrated candidate contains three appliance starts and one day-ahead
heat-pump trajectory.

The controlled whole-home benchmark serializes the same decisions through a
single structured actor interface so that joint numerical scheduling can be
tested without changing the device definitions. Prices, weather, initial
state, appliance data, comfort window, and feeder capacity are frozen before
the first proposal and remain unchanged through critic rounds.

\subsection{Declared physical model}

The day is discretized into $N=96$ consecutive 15-min control steps. Let
$p_t$ denote heat-pump power, $P_t^{\mathrm{tot}}$ whole-home power, $P^{\max}$ the feeder limit,
$\lambda_t$ price, and $T_t$ indoor temperature at state boundary $t$. The
integrated model uses
$P_t^{\mathrm{tot}}=p_t+\sum_aP_ay_{a,t}$, where $P_a$ is appliance power
and $y_{a,t}$ is its reconstructed on/off state; for the heat-pump-only class,
$P_t^{\mathrm{tot}}=p_t$. The critic evaluates
\begin{equation}
\begin{aligned}
J&=0.25\sum_{t=0}^{N-1}\lambda_tP_t^{\mathrm{tot}},\\
T_{t+1}&=a_m(1000p_t)+b_mT_t+c_mT_t^{\mathrm{out}}+d_m,\\
0&\le p_t\le12,\quad T_{t+1}\in[22,24]\ \forall(t+1)\in{\cal C},\\
P_t^{\mathrm{tot}}&\leq P^{\max}\ \quad\text{when the feeder limit is active}.
\end{aligned}
\label{eq:short-model}
\end{equation}
Here $m$ denotes heating or cooling, ${\cal C}$ contains the comfort-state
boundaries, and the factor 1000 converts kW to W. With
$\theta_m=(a_m,b_m,c_m,d_m)$, the thermal model coefficients for heating and cooling are
\begin{equation*}
\begin{aligned}
\theta_{\mathrm{h}}&=(1.5018{\times}10^{-4},.77187,.071008,4.41557),\\
\theta_{\mathrm{c}}&=(-2.2995{\times}10^{-4},.64310,-.005950,9.72241).
\end{aligned}
\end{equation*}
Each published 30-min JEPX price is repeated over two control slots. The feeder
check is active in the integrated MILP evaluation, not in the current
per-device runtime.

\subsection{Fail-closed critic}

The critic first validates the typed representation, reconstructs each binary
cycle, rounds and checks all heat-pump commands, propagates the thermal state
without clipping, and then checks aggregate power when declared. A candidate
that passes is replayed once more without modification before authorization
(Algorithm~\ref{alg:short-policy}). Missing, malformed, failed, or
budget-exhausted candidates are rejected.

\begin{algorithm}[t]
\caption{Fail-closed authorization policy}
\label{alg:short-policy}
\begin{algorithmic}[1]
\REQUIRE Frozen context $w$, candidate $u$, budget $R_{\max}$
\FOR{$r=1,\ldots,R_{\max}$}
  \STATE $v\leftarrow{\cal V}(u;w)$
  \IF{$v$ passes and a fresh ${\cal V}(u;w)$ passes}
    \RETURN \textsc{Authorized}$(u)$
  \ENDIF
  \IF{$r<R_{\max}$}
    \STATE $u\leftarrow\mathrm{ActorRevise}(u,\mathrm{feedback}(v))$
  \ENDIF
\ENDFOR
\RETURN \textsc{Rejected}
\end{algorithmic}
\end{algorithm}

For a violated transition, set
$r_t=b_mT_t+c_mT_t^{\mathrm{out}}+d_m$. Solving the comfort inequality gives
\begin{equation}
\begin{aligned}
q_t^-&=\min\!\left\{\frac{22-r_t}{1000a_m},\frac{24-r_t}{1000a_m}\right\},\\
q_t^+&=\max\!\left\{\frac{22-r_t}{1000a_m},\frac{24-r_t}{1000a_m}\right\},\\
\underline p_t&=\max\{0,q_t^-\},\qquad
\overline p_t=\min\{P^{\max},q_t^+\},\\
{\cal I}_t&=[\underline p_t,\overline p_t]
\quad\text{if }\underline p_t\le\overline p_t.
\end{aligned}
\label{eq:short-interval}
\end{equation}
Ordering the endpoints before intersection makes
Eq.~\eqref{eq:short-interval} sign-safe in both heating and cooling. An empty
interval means that the predecessor state cannot be repaired at that step, so
earlier controls must change or the candidate must be rejected. The initial
policy returned only the first violation. The coupled policy instead supplies
all 40 controls from 08:00--17:45 for a user-specified comfort window, e.g.
09:00--18:00, caps power by residual feeder headroom, identifies conflicting appliances, and requests one complete resubmission. Under the standard occupied-window policy, the critic also rejects any nonzero post-window heat-pump power and requests all controls after the comfort window as one zero-valued batch.
This shutdown is part of the standard coupled policy, not a separate scheduling variant. Deterministic repair is enabled only in the separate heat-pump problem class and is reported independently from actor success.

\subsection{Common acceptance invariant}

Let ${\cal U}$ contain the declared schema, uninterrupted binary-cycle, and
device-bound constraints. Authorization of the exact serialized candidate
$u$ requires a fresh replay satisfying
\begin{equation}
\begin{aligned}
\mathrm{accept}(u)\Rightarrow{}&u\in{\cal U},\\
&T_{t+1}\in[22,24]\quad\forall(t+1)\in{\cal C},\\
&P_t^{\mathrm{tot}}\le P^{\max}\quad\forall t
\quad\text{when declared}.
\end{aligned}
\label{eq:short-invariant}
\end{equation}
Here $P^{\max}=12$~kW for the integrated experiment. Constraints absent from
a problem class are omitted rather than inferred from model behavior. Actor,
deterministic-repair, and solver schedules all enter the same validation path;
therefore a tool call, plausible explanation, or optimizer status alone never
constitutes authorization. In the occupied-window case, replay also requires
$p_t=0$ for post comfort window controls.

\section{Experimental Design}

Three tasks isolate increasing numerical coupling. The MIP task chooses one
valid start for a washing-machine, dishwasher, or EV cycle. The heat-pump LP
uses Eq.~\eqref{eq:short-model} over six reachable and two unreachable frozen
weather-price cases. The whole-home MILP jointly schedules the three binary
cycles and heat pump under a 12-kW feeder limit and a user-specified comfort
window from 09:00--18:00. Pyomo~6.9.3 and Gurobi~11.0
provide matched references \cite{bynum2021pyomo,gurobi2024manual}; solver
outputs pass through the same critic.

For appliance $a$, let $s_a$ be its integer start, $L_a$ its uninterrupted
cycle length, and ${\cal S}_a$ the valid start set induced by release and
deadline constraints. The reconstructed state is
\begin{equation}
y_{a,t}=\mathbf{1}\{s_a\le t<s_a+L_a\},\qquad s_a\in{\cal S}_a.
\label{eq:short-cycle}
\end{equation}
The single-appliance MIP minimizes the tariff-weighted energy of this cycle.
The whole-home MILP combines three such starts with all 96 heat-pump powers,
Eq.~\eqref{eq:short-model}, and
$p_t+\sum_aP_ay_{a,t}\le12$~kW at every slot. The critic reconstructs
Eq.~\eqref{eq:short-cycle} from the serialized starts rather than trusting a
model-supplied on/off array. Thus, an apparently valid start is rejected if
its implied cycle violates a time window or creates a feeder conflict after
integration.

We evaluate local Qwen~3.5 checkpoints at 0.8B, 2B, 4B, 9B, 27B, and 35B-A3B
with temperature-zero decoding and three repetitions. The initial benchmark
contains 252 LLM runs. The current Osaka/Kansai whole-home comparison comprises
three 27B runs, three 35B-A3B runs, and ten matched Pyomo/Gurobi solves using
an August 1, 2026 request, JEPX prices
\cite{jepx2026spot}, Open-Meteo weather \cite{zippenfenig2023openmeteo},
$23^\circ$C initial temperature, and a 22--24$^\circ$C occupied comfort band.
It permits six critic rounds, uses the standard post-window shutdown, and
allows no silent deterministic fallback. Stored artifacts retain every
proposal, validation result, final candidate, token count, and latency.
The other Qwen sizes were not rerun under this final coupled policy and are
therefore excluded from the current coupled comparison. Table~\ref{tab:short-results}
and Fig.~\ref{fig:short-coupled} report only matched current-policy records.

The frozen LP set comprises six reachable heating/cooling cases across Osaka,
Sapporo, and Tokyo and two Osaka cases whose first transition cannot reach the
comfort band from $10^\circ$C or $35^\circ$C. Hourly weather is interpolated
to 15-min resolution. The three binary cases use an eight-slot 2.0-kW washing
cycle, a six-slot 1.8-kW dishwasher cycle, and a 24-slot 7.4-kW EV cycle with
a slot-32 deadline. Reusing the frozen LP snapshots in the original MILP cases
prevents tariff or weather drift across methods.

Initial runs allow four critic evaluations, 1500 output tokens, and a 90-s
generation budget; the updated whole-home runs allow six rounds and 120~s.
Pyomo/Gurobi solves each frozen instance ten times with one thread and zero MIP
gap for integer classes. Agent latency covers proposal, criticism, revision,
and final replay, whereas solver latency includes model construction and
solution; their hardware paths are not normalized. Four post-hoc checks perturb
$a_m$ by $\pm10\%$ or $T_0$ by $\pm2^\circ$C. They measure nominal-schedule
margin under mismatch, not a calibrated uncertainty distribution.

Runs execute sequentially and retain first-load latency. The artifacts record
the serving-software version, exact model digest, request, raw response,
candidate history, critic output, and final serialized schedule. Three
temperature-zero repetitions characterize observed local execution rather than
universal model variance, because serving software and hardware can remain
nondeterministic. A shared-data rule gives the actor and Pyomo the same 96
prices, outdoor temperatures, initial state, coefficients, device data, and
bounds. Solver powers are exported to eight decimal places and then replayed by
the independent critic, which also detects unit, ordering, and serialization
errors outside the optimizer. Regression tests cover price expansion, cost
arithmetic, thermal bounds, unreachable-state rejection, binary windows,
feeder capacity, repair validation, and all three optimization formulations.

\section{Results}

\begin{table*}[t]
\caption{Validated results for the principal problem classes. MIP entries
aggregate 54 single-appliance runs. Coupled entries report the matched
current-policy comparison ($n=3$ per Qwen checkpoint; $n=10$ for
Pyomo/Gurobi). Cost is the mean over accepted schedules only.}
\label{tab:short-results}
\centering
\small
\begin{tabular}{lccccc}
\toprule
Method & MIP feasible & MIP optimal & Coupled accepted & Cost (JPY) & Dominant coupled outcome\\
\midrule
Qwen 3.5 0.8B & 66.7\% & 0.0\% & -- & n/a & timeout\\
Qwen 3.5 2B & 100.0\% & 0.0\% & -- & n/a & timeout\\
Qwen 3.5 4B & 100.0\% & 66.7\% & -- & n/a & timeout\\
Qwen 3.5 9B & 66.7\% & 0.0\% & -- & n/a & power conflict\\
Qwen 3.5 27B & 66.7\% & 66.7\% & \CoupledTwentySevenFeasible & \CoupledTwentySevenCost & accepted\\
Qwen 3.5 35B-A3B & 100.0\% & 66.7\% & \CoupledMoEFeasible & \CoupledMoECost & accepted\\
Pyomo/Gurobi & 100.0\% & 100.0\% & 3/3 & \CoupledGurobiCost & accepted optimum\\
\bottomrule
\end{tabular}
\end{table*}

The single-appliance task achieved \MIPAllFeasible\ feasibility and
\MIPAllOptimal\ exact optimality across 54 Qwen runs
(Table~\ref{tab:short-results}). Continuous coupling was harder: no raw or
critic-revised LP actor trajectory passed the thermal critic, while the
explicit deterministic-repair variant recovered all reachable cases and
rejected both unreachable cases. Under localized feedback, none of the 54
initial whole-home candidates satisfied binary, thermal, and feeder checks
together.

\begin{figure*}[t]
  \centering
  \includegraphics[width=\textwidth]
    {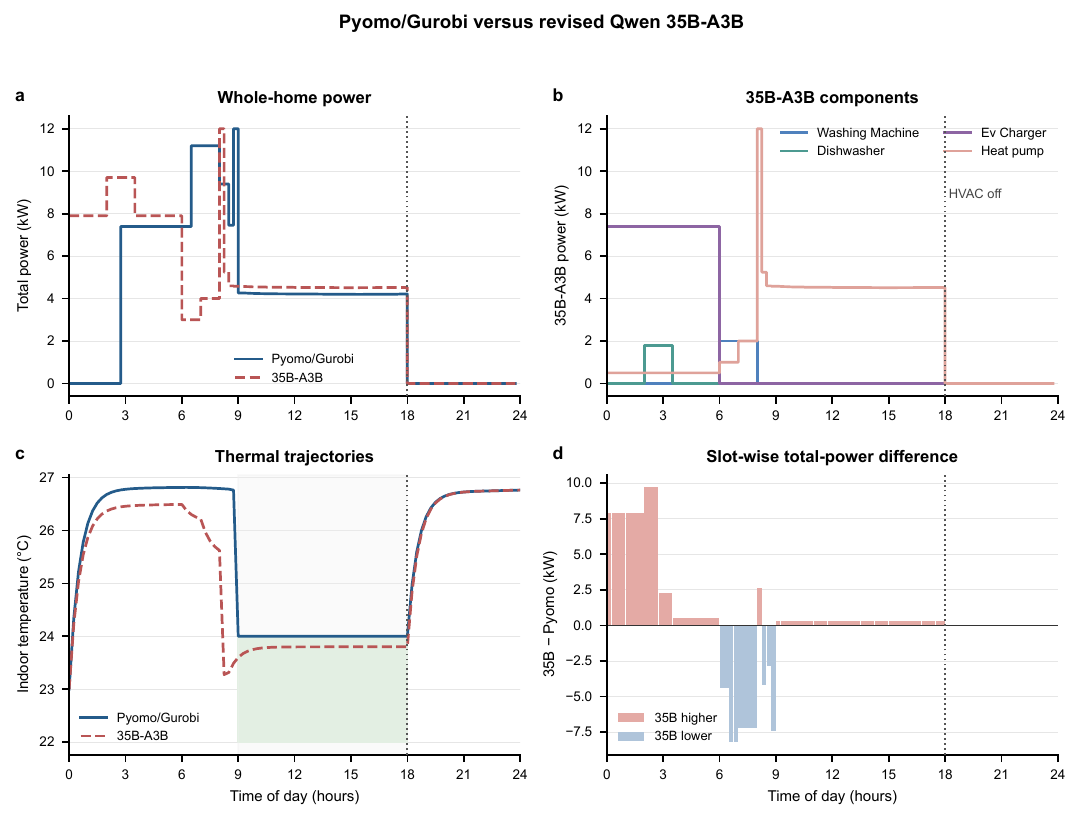}
  \caption{Direct comparison under the standard occupied-window policy.
  (a) Whole-home power for Pyomo/Gurobi and 35B-A3B. (b) Indoor-temperature
  trajectories; the green 22--24$^\circ$C band applies only from
  09:00--18:00. The dotted boundary marks 18:00, after which both heat-pump
  schedules are zero and temperature is unconstrained.}
  \label{fig:short-coupled}
\end{figure*}

Multi-step feedback changed the coupled outcome without relaxing validation.
All \CoupledAllFeasible\ current Qwen schedules were accepted:
\CoupledTwentySevenFeasible\ for 27B and \CoupledMoEFeasible\ for 35B-A3B.
Every accepted schedule satisfied the three uninterrupted cycles, heat-pump
bounds, 12-kW feeder limit, occupied-window comfort constraint, and zero
post-window heat-pump requirement. The three repetitions within each model
were identical. The initial 27B proposal violated the thermal upper bound,
and its complete multi-step revision passed in round two; 35B-A3B passed in
round three.

The 35B-A3B schedule cost \CoupledMoECost~JPY and consumed
\CoupledMoEEnergy~kWh, placing it \CoupledMoEGap\ above the
\CoupledGurobiCost-JPY, \CoupledGurobiEnergy-kWh optimum
(Fig.~\ref{fig:short-coupled}). Its occupied temperature remained
\CoupledMoETempRange$^\circ$C. Pyomo/Gurobi started the washing machine and
dishwasher at slot 26 (06:30) and the EV at slot 11 (02:45); 35B-A3B used
slots 24 (06:00), 8 (02:00), and 0 (00:00). Its heat-pump cost was
843.514~JPY versus 675.531~JPY for the solver.

The 27B schedule cost \CoupledTwentySevenCost~JPY and consumed
\CoupledTwentySevenEnergy~kWh, a \CoupledTwentySevenGap\ gap to the optimum.
It scheduled the dishwasher at 00:00, washing machine at 02:00, and EV at
18:00. The late EV cycle and 87.415-kWh heat-pump trajectory made 27B
\CoupledTwentySevenVsMoECostDelta~JPY
(\CoupledTwentySevenVsMoECostPct) more expensive than 35B-A3B, despite a lower
7.4-kW peak. Both Qwen schedules used
\CoupledHVACEnergyAfterWindow~kWh of heat-pump energy after 18:00; the 27B EV
operation after 18:00 is permitted because the shutdown applies to thermal
control rather than all appliances.

Pyomo/Gurobi required a median \CoupledGurobiMedian~s to build and solve the
case. The actor--critic pipeline averaged \CoupledTwentySevenLatency~s for
27B and \CoupledMoELatency~s for 35B-A3B. These times characterize different
local execution paths and are not a normalized compute benchmark. Because
every trajectory passed the same validator, the reported gaps compare feasible
schedules rather than penalties assigned to invalid outputs.

The experiment does not establish robustness to model error. Post-hoc
coefficient and initial-state perturbations reduced feasibility, showing that
Eq.~\eqref{eq:short-invariant} is conditional on the encoded coefficients and
state. Calibrated uncertainty sets, tightened margins, or receding-horizon
measurement feedback are required before broader operational claims.

\subsection{Safeguarding and failure interpretation}

The heat-pump LP separates numerical feasibility from actor behavior. Gurobi
returned a critic-valid optimum for every reachable case and declared both
extreme-initial-state cases infeasible. No reachable raw or critic-revised
actor trajectory passed the thermal replay, so all 108 reachable LP runs used
the explicitly labeled deterministic-repair path. Repair selected commands
only from nonempty feasible intervals and revalidated the complete horizon;
unreachable cases preserved rejection rather than returning a completed but
physically invalid array. Consequently, deterministic safeguarding achieved
feasible final outputs without being counted as unaided LLM scheduling.

The recorded failures occurred at different stages. A response could omit the
critic call, provide a malformed array, exceed the time limit, respect the pointwise power range while
violating a later thermal transition, fail to replace the complete trajectory
after feedback, retain a coupled feeder conflict, or exhaust its API budget.
Only some of these are thermal-reasoning failures. Schema adherence, tool-call
syntax, and endpoint availability are distinct from physical feasibility, so
tool use and checkpoint scale cannot serve as authorization proxies. The
current coupled results illustrate this distinction: 27B passed after two
critic rounds but was slower and \CoupledTwentySevenVsMoECostPct\ more
expensive than 35B-A3B, which passed after three rounds.

Mathematical programming therefore serves as a numerical reference rather
than a replacement for the agentic layer. The frozen LP has affine dynamics
and a linear objective, while appliance cycles make the integrated task a
compact MILP. Pyomo/Gurobi supplies an optimum or infeasible status for these
declared models; the agentic workflow supplies intent interpretation,
specialist delegation, revision, and an auditable explanation of rejection.
Both routes terminate at the same independent critic. A future implementation
may expose an optimizer as another specialist tool, but its serialized output
must still satisfy Eq.~\eqref{eq:short-invariant} before execution.

\subsection{Operational scope}

The authorization claim is deliberately narrower than deployment safety. The
critic establishes that an accepted day-ahead schedule satisfies the encoded
schema, uninterrupted appliance cycles, device power bounds, surrogate thermal
dynamics, occupied-window comfort band, and the shared feeder limit when that
constraint is declared. It does not validate omitted dynamics or infer missing
constraints from an LLM explanation. In particular, the present case does not
model humidity, heat-pump cycling and ramping, battery state of charge,
appliance interruption, forecast uncertainty, protective relays, communication
loss, or occupant overrides. The linear coefficients are prototype surrogates,
and the weather-price cases are controlled evaluations rather than a
representative sample of Japanese homes.

Operational use would therefore require identified uncertainty sets,
equipment-specific constraints, and receding-horizon replay against measured
states. A deployment-oriented critic could tighten the comfort and feeder
margins, invalidate stale forecasts, and reject a previously authorized plan
when telemetry diverges from the frozen context. These additions would preserve
the same fail-closed boundary: LLM-based specialists may interpret intent and propose
revisions, while only a deterministic evaluation of the exact action object may
release a schedule. Field testing and a multi-agent-versus-monolithic ablation
remain outside the current evidence.

\section{Conclusion}

This paper presents a hierarchical agentic HEMS with a deterministic physical
authorization boundary. The decisive result is that coupled multi-step critic
feedback enabled all \CoupledAllFeasible\ current 27B and 35B-A3B schedules
while every invalid candidate remained blocked. Under the standard
occupied-window policy, 35B-A3B reached a \CoupledMoEGap\ cost gap to the
matched optimum and both Qwen models used zero heat-pump energy after 18:00.
The framework therefore supports natural-language
coordination and critic-validated MIP/MILP scheduling under one frozen model.
Acceptance remains conditional on the encoded schema, thermal surrogate,
device limits, and integrated feeder constraint; humidity, equipment cycling,
battery state of charge, calibrated uncertainty, electrical protection, and
closed-loop field measurements are outside the present model.

\bibliographystyle{IEEEtran}
\bibliography{ref}

@article{elmakroum2026agentic,
  author  = {Reda El Makroum and Sebastian Zwickl-Bernhard and Lukas Kranzl},
  title   = {Agentic {AI} Home Energy Management System: A Large Language Model Framework for Residential Load Scheduling},
  journal = {Results in Engineering},
  volume  = {29},
  pages   = {109857},
  year    = {2026},
  doi     = {10.1016/j.rineng.2026.109857}
}

@article{jia2025feedback,
  author  = {Mengshuo Jia and Zeyu Cui and Gabriela Hug},
  title   = {Enhancing {LLMs} for Power System Simulations: A Feedback-Driven Multi-Agent Framework},
  journal = {IEEE Transactions on Smart Grid},
  volume  = {16},
  number  = {6},
  pages   = {5556--5572},
  year    = {2025},
  doi     = {10.1109/TSG.2025.3589114}
}

@article{jiang2026physics,
  author  = {Zixin Jiang and Weili Xu and Bing Dong},
  title   = {An Agentic {AI}-Enabled Physics-Informed Machine Learning Framework for Grid-Interactive, Decarbonized Building Operations},
  journal = {Advances in Applied Energy},
  volume  = {22},
  pages   = {100273},
  year    = {2026},
  doi     = {10.1016/j.adapen.2026.100273}
}

@article{jung2026hema,
  author  = {Wooyoung Jung},
  title   = {Multi-Agent Home Energy Management Assistant ({HEMA})},
  journal = {SoftwareX},
  volume  = {34},
  pages   = {102633},
  year    = {2026},
  doi     = {10.1016/j.softx.2026.102633}
}

@inproceedings{michelon2025interface,
  author    = {Fran{\c{c}}ois Michelon and Yihong Zhou and Thomas Morstyn},
  title     = {Large Language Model Interface for Home Energy Management Systems},
  booktitle = {Proceedings of the 16th ACM International Conference on Future and Sustainable Energy Systems},
  pages     = {590--602},
  year      = {2025},
  doi       = {10.1145/3679240.3734586}
}

@inproceedings{yao2023react,
  author    = {Shunyu Yao and Jeffrey Zhao and Dian Yu and Nan Du and Izhak Shafran and Karthik Narasimhan and Yuan Cao},
  title     = {{ReAct}: Synergizing Reasoning and Acting in Language Models},
  booktitle = {The Eleventh International Conference on Learning Representations},
  year      = {2023},
  url       = {https://openreview.net/forum?id=WE_vluYUL-X}
}

@article{bonabeau2002agent,
  author  = {Eric Bonabeau},
  title   = {Agent-Based Modeling: Methods and Techniques for Simulating Human Systems},
  journal = {Proceedings of the National Academy of Sciences},
  volume  = {99},
  number  = {suppl. 3},
  pages   = {7280--7287},
  year    = {2002},
  doi     = {10.1073/pnas.082080899}
}

@misc{jepx2026spot,
  author       = {{Japan Electric Power Exchange}},
  title        = {Spot Market Data},
  year         = {2026},
  howpublished = {\url{https://www.jepx.jp/electricpower/market-data/spot/}},
  note         = {Accessed: 2026-07-27}
}

@misc{zippenfenig2023openmeteo,
  author    = {Patrick Zippenfenig},
  title     = {Open-Meteo.com Weather {API}},
  year      = {2023},
  publisher = {Zenodo},
  doi       = {10.5281/zenodo.7970649}
}

@book{bynum2021pyomo,
  author    = {Michael L. Bynum and Gabriel A. Hackebeil and William E. Hart and Carl D. Laird and Bethany L. Nicholson and John D. Siirola and Jean-Paul Watson and David L. Woodruff},
  title     = {Pyomo---Optimization Modeling in Python},
  edition   = {3},
  publisher = {Springer},
  address   = {Cham, Switzerland},
  year      = {2021},
  doi       = {10.1007/978-3-030-68928-5}
}

@manual{gurobi2024manual,
  author       = {{Gurobi Optimization, LLC}},
  title        = {Gurobi Optimizer Reference Manual, Version 11.0},
  organization = {Gurobi Optimization, LLC},
  year         = {2024},
  url          = {https://docs.gurobi.com/_/downloads/optimizer/en/11.0/pdf/}
}

\end{document}